\documentclass[
]{ceurart}

\usepackage{listings}
\usepackage{caption}
\begin{document}

\copyrightyear{2024}
\copyrightclause{Copyright for this paper by its authors.
  Use permitted under Creative Commons License Attribution 4.0
  International (CC BY 4.0).}

\conference{Scholarly QALD challenge at ISWC 2024}

\title{Contri(e)ve: Context + Retrieve for Scholarly Question Answering}

\author[1]{Kanchan Shivashankar}[%
orcid=0009-0004-4948-9502,
email=shivashankar@uni-wuppertal.de,
]
\address[1]{University of Wuppertal, Germany}

\author[2]{Nadine Steinmetz}[%
orcid=,
email=nadine.steinmetz@fh-erfurt.de,]
\address[2]{University of Applied Sciences Erfurt}

\begin{abstract}
Scholarly communication is a rapid growing field containing a wealth of knowledge. However, due to its unstructured and document format, it is challenging to extract useful information from them through conventional document retrieval methods. Scholarly knowledge graphs solve this problem, by representing the documents in a semantic network, providing, hidden insights, summaries and ease of accessibility through queries. Naturally, question answering for scholarly graphs expands the accessibility to a wider audience. But some of the knowledge in this domain is still presented as unstructured text, thus requiring a hybrid solution for question answering systems. In this paper, we present a two step solution using open source Large Language Model(LLM): Llama3.1 for Scholarly-QALD dataset. Firstly, we extract the context pertaining to the question from different structured and unstructured data sources: DBLP, SemOpenAlex knowledge graphs and Wikipedia text. Secondly, we implement prompt engineering to improve the information retrieval performance of the LLM. Our approach achieved an F1 score of 40\% and also observed some anomalous responses from the LLM, that are discussed in the final part of the paper.
\end{abstract}


\maketitle

\section{Introduction}
Scholarly knowledge is mostly represented as scholarly articles and growing rapidly. These documents contain invaluable knowledge that cannot be easily automated for extraction and hence not machine actionable. They contain a heterogeneous source of structured and unstructured data. It is important to study them as they contain important information critical for further advancements in academia and a network of unexplored relationships. Due to its unstructured nature, conventional information and document retrieval processes, albeit efficient, are limited and fail to provide new connections and insights. Scholarly knowledge graphs are necessary to fill this gap and represent this knowledge in a structured and machine readable format. This has led to new scholarly knowledge graphs such as DBLP, OpenAlex, ORKG etc.

Knowledge graphs capture the semantic representation of the underlying data, by defining meaningful concepts and relationships. This enriches the knowledge and established relationships between concepts. The structured representation can lead to easier access by setting up knowledge graph question answering systems. This paper details our contribution for participating in the second Scholarly-QALD challenge as part of ISWC-2024. The challenge is to find a solution for a hybrid question answering system for scholarly data. The answer to the question must be extracted from the three knowledge sources, which include, two scholarly knowledge graphs: DBLP and SemOpenAlex and an unstructured source: Wikipedia text. 

In the upcoming sections of the paper, we first look at the relevant literature to provide background information to the reader, followed by the methodology section that provides a detailed account of our contribution. Subsequently, we present the results and discuss some of the gaps identified in the dataset and knowledge sources, before providing a conclusion.

\section{Related Work}
Scholarly knowledge graph provides a semantically connected graph for scientific and academic literature. With the rapid growth of scholarly data storing it as linked data facilitates transformation of document format to a machine understandable and actionable data. The work in\cite{verma_scholarly_2023} reviews three different ways: machine learning, rule-based learning and natural language processing tools for constructing a scholarly knowledge graph. It also highlights some of the existing challenges such as integration of data sources, ontology matching, extracting KG from diversely structured textual data etc. To fill the existing gap in knowledge the work in\cite{jia_leveraging_2024} proposes a novel approach to enhance the construction of scholarly KGs. The two innovative methods introduced provides a structured representation of document-formatted scholarly text(Deep Document Model) and a query processor to optimize and simplify complex queries (KG-enhanced Query Processing). Automation of KG construction has led to many new and popular scholarly KGs such as DBLP, OpenAlex, ORKG etc.

DBLP stands for Data Bases and Logic Programming. Created in 1993, at University of Trier, it was originally designed as a database to store bibliographic information for the fields of database systems and logic programming. Today, it extends to a much wider range of sub-fields of computer science\cite{banerjee_dblp-quad_2023}. SemOpenAlex knowledge graph contains over 26 billion RDF triples covering all areas of scientific scholarly domain. It overcomes the limitations of some of the existing scholarly KG, such as, limited focus to a specific discipline, not regularly updated or does not follow standard RDF format\cite{farber_semopenalex_2023}. ORKG (Open Research Knowledge Graph) proposes a transition from document based scholarly communication to a knowledge-based machine actionable format. The creation of this knowledge can be done by the users themselves through a front-end(also includes search options) and the request is processed in the back-end by validating the logic\cite{jaradeh_open_2019-1}.

With the establishment of scholarly KG, one of its application is Scientific Question Answering. JarvisQA\cite{jaradeh_question_2020} proposes a BERT-based question answering system that retrieves answers from different tables. Another application is author disambiguation in scholarly KG. This is an important challenge due to the authors central link in the KG and existence of multiple authors with similar name, field and affiliation. A triangulation approach is proposed to improve evaluation of author name disambiguation in DBLP \cite{kim_evaluating_2018}. The improved evaluation is attributed to assessing the performance by comparing baselines on eight labeled datasets.

Generative AI tools like large language models are providing peak performance in natural language generation. However, some of their biggest challenges are knowledge gap from lack of training data and hallucinations. They have greatly improved from efforts of tweaking the prompt to force a better response. Knowledge Injection i.e, providing contextual information in the prompt about the relevant entities, reduced hallucinations and improved the quality of responses as one case study in text generation for online customer reviews reports\cite{Martino_KI_2023}. 

Using LLM to solve KGQA have also become popular. For example, the paper\cite{taffa2023leveraging} leverages LLM to generate SPARQL query generation using zero-shot and few-shot prompts. In the few-shot prompt setting a BERT-based question analyser identifies similar question-query pairs from the dataset and introduces them into the prompt. The model achieved F1-score of 99\% on the Sci-QA dataset. A use-case in virology showcases the power of LLMs in extracting scientific information with the help of instruction fine-tuning\cite{shamsabadi2024large}. A filtered corpus of CORD-19 dataset(papers published on COVID-19) is used to fine-tune an LLM for better domain adaptation. LLMs have proven their ability of handling both structured and unstructured data. This gave us the idea to use LLMs to build our hybrid QA system. 

\section{Methodology}
This section discusses our contribution in this paper. We have a two step approach of context extraction and prompt engineering. Before discussing our work, we first take a look at the question answering dataset.

\subsection{Dataset}
The scholarly-QALD dataset is the second iteration of Question Answering dataset for scholarly data, preceded by DBLP-QuAD dataset\cite{banerjee_dblp-quad_2023}. The dataset is split into train and test set and  was generated using an LLM, which introduces a small percentage of incorrect answers to the dataset. The train dataset consists of 5000 question-answer pairs with each question containing DBLP author-id for the question. The test set consists of 702 questions with DBLP author-id. The answers are held-out and to mitigate incorrect answers due to LLM generation, they have also been manually verified and corrected.

This is a hybrid question-answering challenge i.e, the correct answer to the question requires the system to look up three different sources of data (DBLP KG, SemOpenAlex KG and Wikipedia text). DBLP KG is limited to representing only the computer science scholarly communication, whereas, SemOpenAlex is an open source KG representing all scientific disciplines. SPARQL endpoints are provided to access the two knowledge graphs. The Wikipedia source file contains text related to authors and institutions, extracted from its respective Wikipedia page.

\subsection{Context Extraction}
\begin{figure}
    \centering
    \includegraphics[width=0.7\linewidth]{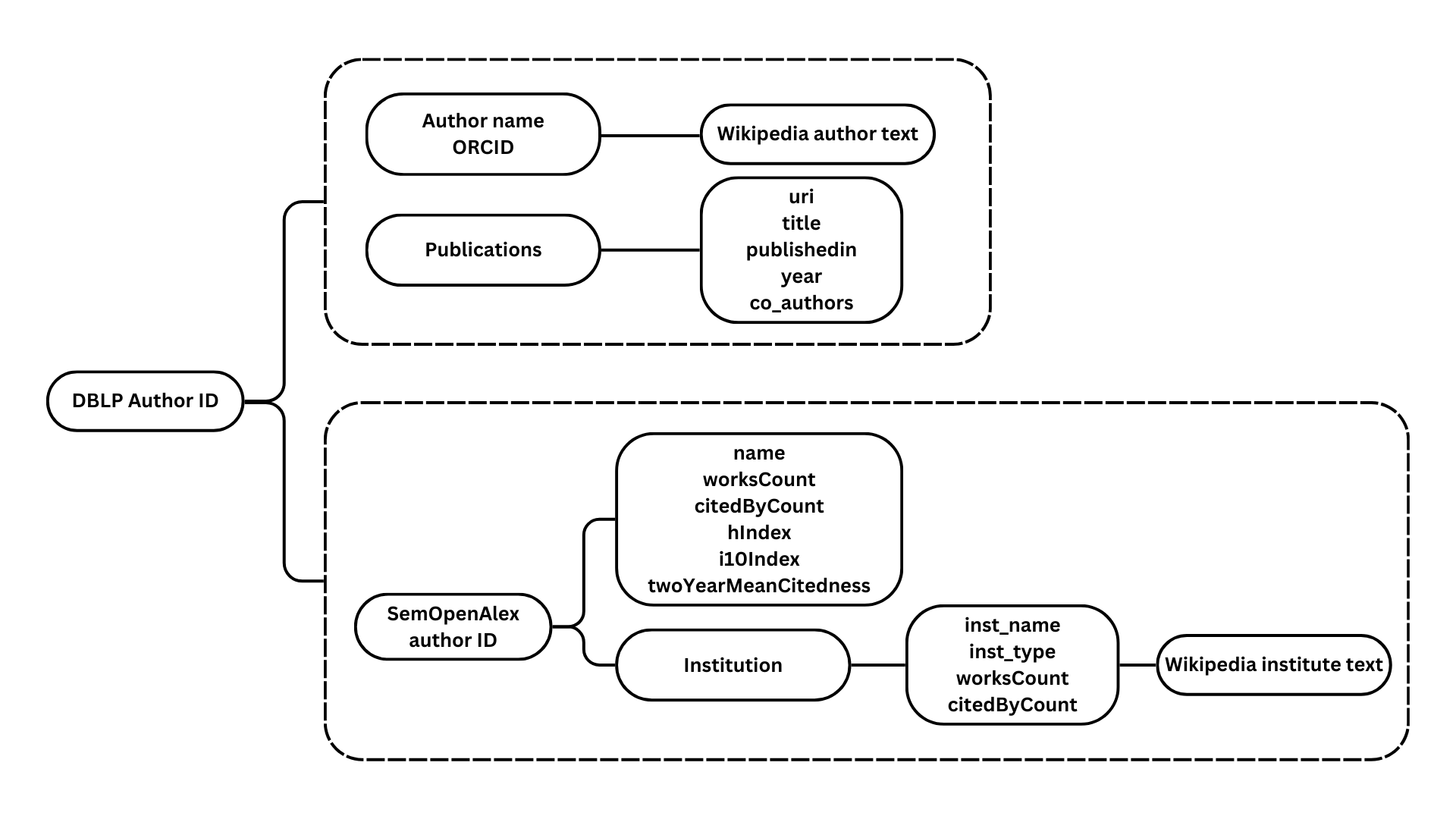}
    \caption{Connectivity between the data sources}
    \label{fig:data_flow_chart}
\end{figure}

The Scholarly-QALD dataset was introduced in the previous section, containing question and DBLP author-id in the form of URIs,(Uniform Resource Identifiers) required to answer each question. The DBLP author-id can be used to connect to all the other sources, as demonstrated in figure\ref{fig:data_flow_chart}.

Before we proceed with answering the question, we need to gather all the relevant information about the author, on whom the question is formed. This step is used to extract the relevant information from the different sources. The process of extracting context is divided into three steps

\begin{enumerate}
    \item Use DBLP author-id to fetch author information(including ORCID) from DBLP KG
    \item Using DBLP ORCID to extract author information from SemOpenAlex KG
    \item Extract author and affiliated institution text from Wikipedia text
\end{enumerate}

\subsubsection*{DBLP Knowledge Graph}\footnote{https://dblp.org/}
DBLP author-id corresponding to each question is provided in the dataset. Author-id URI links all the information related to a particular author in DBLP. By studying the questions in the dataset we found that only some of these properties are relevant for answering the question. The SPARQL query was refined to extract ORCID, name and publication information of the author.

An author can have more than one publication. Title of the publication, published journal/series/book name, publication year and co-authors were extracted for each of the author's publication to provide as context.

\subsubsection*{SemOpenAlex Knowledge Graph}
The ORCID extracted from DBLP knowledge graph in the previous step, is used to fetch the SemOpenAlex author-id. This author-id gives us all the information related to the author and their affiliated institution. The SPARQL query extracts author properties such as, author name, publication count, citations count, hIndex, i10Index, two year mean citedness and author affiliated institution name. Additionally, we also extract institution type, citation count and publication count of the author affiliated institution.

\subsubsection*{Wikipedia}
Some questions relate to author/institute that is not represented in the knowledge graphs. The Wikipedia text file provided contains information for only some of the authors and institutions. It is stored in a dictionary-like format with names and description as key-value pair. The author and their affiliation names, extracted from the two KGs, are used to do a fuzzy search in the file. If an entry exists in the file, the description text is extracted, else, we extract the content from the Wikipedia page directly.

These texts are lengthy and providing them directly as context in the prompt resulted in poor performance from the LLM. To mitigate this issue, we extracted keywords from the question and did a substring match with the text. Only sentences containing the keyword is included as the final text to be provided in the prompt. Summarizing the Wikipedia text improved the performance greatly.

After the conclusion of the three step information extraction, the data is further refined and combined to provide pertinent information as context in the prompt. Details are discussed in the next section. This step was necessary, as longer prompt and context length resulted in poor performance by the LLM, both with execution times and accuracy.

\subsection{Prompt Engineering}
The prompt template is designed to provide all the relevant information for the LLM to infer the correct answer.

As discussed earlier, the length of prompt determines the performance of the LLM. If the prompt length is too short there might not be enough information to extract the answer, resulting in hallucinations. Longer prompts increases the LLM inference time significantly and frequently outputs wrong answers. Thus, it was necessary to further refine the prompt.

The prompt contains 4 parts,
\begin{itemize}
    \item Instructions: guides the LLM on how to process the information in the prompt
    \item Query: question from the dataset and a rephrased form of the original question generated by the LLM
    \item Context: information containing the answer extracted from the three sources in the previous step. It is grouped into Author, Institute and Publication
    \item Output Indicator: instructions to generate the output, which also includes the answer type predicted for each question\cite{shivashankar2021reaching}
\end{itemize}

The prompt template is provided below.

\begin{verbatim}
    #Instructions:
    You are a factual question answering machine about scholarly data.                                              
    Answer the Question. The answer is present within the context.
    The Context information about the author is provided in three parts with 
    (author/institute, property, value) triples and text (1) Author information 
    (2) Affiliated institution information (3) Publications information.
    Make use of all the context information available to answer the question.
    
    #Query: 
    Question: {question}
    Simplified Question: {rephrased_question}
    
    #Context:
    1 Author: 
        {semoa_auth_data} #SemOpenAlex author information
        {wiki_auth_context} #Wikipedia author text
    2 Affiliated Institute:
        {semoa_inst_data} #SemOpenAlex institute information 
        {wiki_inst_context} #Wikipedia institute information
    3 Publications:
        {dblp_pub_data} #DBLP publication information
        
    #Output Indicator:
    Return only the answer value. Do not clarify or explain.
    Answer must match the answer type provided.
    Answer type format is number: (float, integer), date(dd MM yyyy or yyyy), 
    string, or resource(string)
    Answer type: {type}
    Answer must be a name and not url.
    Answer "Do not know" if you are not sure.
\end{verbatim}

We used open source llama3.1 8b-Instruct model\cite{dubey2024llama3herdmodels} released from Meta AI(July 2024), for building our hybrid QA system. The model is trained on 8 billion parameters and fine-tuned with natural language instructions. To avoid memorization of the dataset and answers by the LLM, the prompts were input each time in a random order. The output of the prompt is recorded as the final result. You can find more details on this in the results section of the paper.

\section{Results}
The table below provides the final evaluation scores of our approach for Scholarly-QALD test dataset. 

The two main evaluation measures used in the challenge are Exact Match and F1-score. Exact Match gives the total percentage of generated answers that match accurately with the golden answers. Since the answers in the dataset are LLM generated, F1-score is used to compute a word-to-word match score for partially correct answers.

\begin{table}
    \begin{center}    
    \caption{Scholarly-QALD Evaluation Scores}
    \label{tab:my_label}
    \begin{tabular}{|c|c|c|}
        \hline
        Dataset & Test\\
        \hline
        Exact Match Accuracy  & 32.05\% \\
        \hline
        Precision  & 0.40917 \\
        \hline
        Recall & 0.442 \\
        \hline
        F1 & 0.40702 \\
        \hline
    \end{tabular}
    \end{center}
\end{table}

Our system achieved the second best results in the challenge(out of 3). However, the low scores necessitates a further inspection. In the following section, we discuss the results generated by LLMs.

\subsection{Analysis}
\subsubsection*{Missing author ORCIDs}
As discussed earlier ORCID serves as unqiue id for each author, that forms a common link for connecting authors between DBLP and SemOpenAlex knowledge graphs. However, we found that many ORCIDs were missing in SemOpenAlex knowledge graph(in the SPARQL endpoint provided for the challenge), which led to incomplete context extraction. To mitigate this issue we accessed a newer version of the knowledge graph using a different SPARQL endpoint\footnote{https://semopenalex.org/sparql}, but this also had updated values(such as publication count, h-Index values etc.). This has potentially led to some mismatches in the systems answers.

\begin{verbatim}
Id: 9a2a668a-89e7-4506-857f-b301bd1d2074
Question: What is the institute where the author of Built-in generation of 
functional broadside tests graduated from?
Dblp_author_id: <https://dblp.org/pid/p/IrithPomeranz>
Orcid: <https://orcid.org/0000-0002-5491-7282>
\end{verbatim}

\subsubsection*{LLM inconsistencies}
In spite of the answer being present in the context the responses produced by the LLM was ambiguous and inconsistent. These behaviors were random and did not seem to follow a pattern. Such cases are highlighted below

\begin{itemize}
    \item The LLM showed a random behavior of hallucinating values in the answer, although the correct value was provided as part of the context. This was especially common with number answer types.
    \begin{verbatim}
        Id: b00d5b01-c0a1-4e5e-be22-46a58df50417
        Question: In which year did the author who contributed to the 
        Replacement - A Sheffer Stroke for Belief Change work become a 
        professor at KTH?
        Gold answer: 2000
        System answer: 1951
    \end{verbatim}
    
    \item The LLM generated gold answers observed in the train dataset did not follow a fixed pattern. For example, floating point precision after the decimal points, date format etc. This resulted in a few mismatches with the golden answers. Listed below are some of the examples from train dataset, for which the gold answers were provided.
    \begin{verbatim}
        Example 1:
        Id: 48f41d21-b0fb-45b7-a6e4-96495160f2d7
        Question: What is the average two years citedness of the writer of 
        Fundamental Analysis of Lateral Displacement Estimation Quality in 
        Ultrasound Elastography?
        Gold answer: 3.4050634
        System answer: 3.4050633907318115

        Example 2:
        Id: 8810c40f-0194-4bc3-bb2a-caa668d343be
        Question: What is the birth date of the writer of The Impact of 
        Robotics on Computer Science?
        Gold answer: Oct 7, 1939
        System answer: 07 10 1939
    \end{verbatim}
    
    \item The LLM had a hard time generating only the answer keywords, ignoring the specific instructions provided in the prompt. Which led to full sentences or long explanations provided along with the answer.
    \begin{verbatim}
        Example 1:
        Id: 9bfb08b8-c06e-4d09-971c-726ff44eaa77
        Question: Where did the contributor of Alternative Digit Sets for 
        Nonadjacent Representations receive his B.Math degree from?
        System answer: The University of Waterloo is located in Waterloo, 
        Ontario, Canada.

        Example 2:
        Id: 5bc75b81-7716-40d6-87f2-e92480d3d37e
        Question: For what outstanding contribution did the author of 
        Enabling Technology for On-Chip Interconnection Networks receive 
        the 2010 ACM/IEEE Eckertâ-Mauchly Award?
        System answer: The 2010 ACM\/IEEE Eckert\u2013Mauchly Award was 
        received by the author of \"Enabling Technology for On-Chip 
        Interconnection Networks\" for their work on \"the development of 
        high-speed interconnects and networks-on-chip.\"
        Therefore, the answer is: (enabling technology for on-chip 
        interconnection networks)
    \end{verbatim}
\end{itemize}

\section{Conclusion}
In this paper, we proposed a context in prompt solution using Llama 3.1, to solve hybrid question answering problem in the scholarly domain. The hybrid data sources included DBLP and SemOpenAlex knowledge graphs and Wikipedia text. The two step solution extracted and input the context into prompts, to evaluate the information retrieval capabilities of LLM. The results showed potential, but the performance of the LLM was inconsistent. This behavior dictates a need for further study of attention paid to context in prompts and to understand the information retrieval capabilities of LLMs.

\bibliography{sample-ceur}

\end{document}